\documentclass[onecolumn,12pt]{article}
\usepackage{amssymb}
\usepackage{graphicx}
\usepackage{graphics}
\usepackage{enumerate}
\usepackage{geometry}

\newcommand\ovl[1]{\overline{#1}}
\newcommand\udl[1]{\underline{#1}}

\newcommand {\ket}[1] {|{#1}\rangle}

\newcommand {\bra}[1] {\langle{#1}|}

\newcommand {\cl}{\mathcal}
\newcommand {\tsf} [1]{\textsf{#1}}
\newcommand {\norm} [1] {\parallel #1 \parallel}
\newcommand {\al} {\alpha}
\newcommand {\e} {\epsilon}
\newcommand{\tr} {\tsf{tr}}
\newcommand {\beq} {\begin{equation}}
\newcommand {\eeq} {\end{equation}}
\newcommand {\mrm}{\mathrm}

\begin{document}

\title{Classicalization of Nonclassical Quantum States in Loss and Noise -- Some No-Go Theorems}
\author{Horace P.~Yuen$^{1,2}$\footnote{Email: yuen@eecs.northwestern.edu}\hspace{2mm} and Ranjith Nair$^1$ \\ $^1$Department of Electrical Engineering and Computer Science\\
$^2$Department of Physics and Astronomy\\
Northwestern University, Evanston, IL 60208} \maketitle

\newline
\begin{abstract}
The general problem of performance advantage obtainable by the use of nonclassical transmitted states over classical ones is considered. Attention is focused on the situation where system loss is significant and additive Gaussian noise may be present at the receiver. Under the assumption that the total received state is classical, rigorous output density operator representations and their trace distance bounds are developed for classical and nonclassical transmitted states. For applications with high loss in all modes, a  practical No-Go theorem is enunciated that rules out the possibility of significant advantage of nonclassical over classical states. The recent work on quantum illumination is discussed as an example of our no-go approach.
\end{abstract}

\section{Introduction}

Electromagnetic fields are widely used for communication, sensing, and precision measurement including the ongoing efforts on the detection of gravitational radiation. The quantum states of the electromagnetic fields that are commonly produced from conventional sources are either the coherent states or their random mixtures, which constitute the class of \emph{classical states} of the radiation fields. It has long been known that states outside this class, the \emph{nonclassical states} such as squeezed states \cite{yuen76,yuen04} and photon number eigenstates, could lead to dramatic performance improvement in the ideal limit for the above mentioned applications. This happens for just a single space-time field mode with a given average energy, so that the total excited field is still fully space-time coherent. On the other hand, it is also well known \cite{yuen76,yuen04} that noise and especially loss puts a serious limit on the advantage of nonclassical states compared to classical ones.

Recently, the effects of quantum entanglement have aroused a great deal of interest in the  area of quantum information. An entangled quantum state of the radiation field is necessarily nonclassical. It offers the possibility that one component of the two entangled subsystems may avoid, at least in the ideal limit, the action of noise and loss by being retained locally, while the other component would be sent out to the environment with loss necessarily introduced into it, and perhaps noise also. A prime example is a (monostatic) radar system in which one component of the entangled state is transmitted for target detection. A main question is: Can such an entangled system provide significant advantage when only the transmitted mode is significantly corrupted by loss and noise?

The answer to this question is clearly quantitative and also depends on the specific performance criterion under consideration. If the loss and noise is very small, we would expect large advantage comparable to the ideal lossless and noiseless limit, and if they are sufficiently big, we would expect no meaningful advantage. In this paper, we try to quantify the answer generally by comparing the density operator of the total received state from a nonclassical transmitted state to that obtained from a classical transmitted state of comparable energy. For the case of one or two field modes, and when the total \emph{received} state is classical, we give explicit trace distance bounds between these two density operators, which in turn quantify the performance difference for any criterion of interest. In the general multimode case, we delineate conditions under which the two density operators described above fall into the same equivalence class, in practice at least, so that a \emph{Practical No-Go} result is obtained which indicates no meaningful performance advantage can be obtained from the use of entangled and also nonclassical states in general -- this No-Go applies when the transmittance $\kappa \ll 1$ in \emph{all} the modes.

The results of Section IV are obtained under the condition that the total received state is classical as a result of noise added to the nonclassical transmitted state. Note that a received state that is classical from channel loss and noise does \emph{not} imply that it can be obtained from a classical transmitted state of the same average energy. A striking example is the case of ``quantum illumination'' discovered recently \cite{lloyd08,tan08}, where our general No-Go approach applies, even though the loss is not large in the modes held at the transmitter.  A practical no-go obtains at optical frequencies and below, as detailed in Section V, since one photon per mode translates into a small power at such frequencies. Except in the case of high loss in all modes to which our practical no-go applies, the \emph{general} question of advantage obtainable from the use of nonclassical states is open -- viz., there is no definite rigorous result that would show whether or not significant advantage can be obtained.

In Section II we review some basic facts of particular relevance from quantum optics. The Classicalization Theorem, which indicates how close to classical any quantum state becomes in the presence of loss and noise is established in Section III. In Section IV rigorous trace distance bounds for the receiver states obtained from classical versus nonclassical transmitted states are derived. In Section V general arguments are presented that identify such receiver states as ``equivalent'' at least in practice, and the quantum illumination examples are discussed. Concluding remarks are given in Section VI.

\section{Classical versus Nonclassical States}

For $m$ bosonic modes described by a tensor product
$\cl{H}= \bigotimes_{i=1}^m \cl{H}_i$ of infinite-dimensional
Hilbert spaces $\cl{H}_i$, let $\ket{\underline{\alpha}}=\prod_i
\ket{\alpha_i}$ be the coherent states of the modes, with
$a_i\ket{\alpha_i}= \alpha_i\ket{\alpha_i}$ for the $i$th modal
annihilation operator $a_i$. A joint state $\rho$ on $\cl{H}$ is a
\emph{classical state} if it can be represented as
\begin{equation} \label{classical}
\rho = \int P(\udl{\alpha})\ket{\udl{\alpha}}\bra{\udl{\al}}
\textrm{d}^2 \udl{\alpha}
\end{equation}
where $P(\udl{\al})$ is a \emph{true} probability density function. Otherwise, the state is \emph{nonclassical}. A classical state is necessarily un-entangled because (\ref{classical}) provides a
separable form for it. However, the marginal states, say the
state of the signal or idler mode -- $\rho_s = \tr_i \rho_{si}$ or
$\rho_i = \tr_s \rho_{si}$ -- of an entangled (hence
nonclassical) state may be classical, as is the case for the output of a downconversion process.

The following basic facts of quantum optics may be recalled, which are of course applicable to any frequency. A quantum state $\rho$ is completely characterized by its characteristic functions. The three common ones are the normally ordered, anti-normally ordered, and symmetrically ordered (Weyl) characteristic functions, which for a single mode are defined as:
\begin{eqnarray}  \label{charfuncs} \chi_N(\mu) &=& \tr \rho e^{-\mu
a^{\dag}}e^{\mu^{\ast}a}, \nonumber\\ \chi_A(\mu) &=& \tr \rho
e^{\mu^{\ast}a} e^{-\mu a^{\dag}},\\ \chi_W(\mu) &=& \tr \rho
e^{-\mu a^{\dag}+\mu^{\ast}a}. \nonumber \end{eqnarray}
 They are related by

\beq \label{charrel} \chi_N(\mu) =
e^{|\mu|^2/2} \chi_W(\mu) = e^{|\mu|^2} \chi_A(\mu). \eeq

The Fourier transform of
$\chi_W(\mu)$ is the Wigner distribution $W(\al)$, that of
$\chi_A(\mu)$ is the Q-function $Q(\al) \equiv
\bra{\al}\rho\ket{\al}/{\pi}$ -- both exist in the usual
$L^1$-sense. The Fourier transform of $\chi_N(\mu)$ is the
$P(\al)$ of (\ref{classical}), and exists only in a subtle
distributional sense. While $Q(\al)$ is a smooth true probability
density, $W(\al)$ is smooth but can take negative values. All these notions generalize immediately to the multimode case for a canonical set of modal operators $\{a_i\}$ with $[a_i,a_j^{\dag}]=\delta_{ij}$.

We have the following important result from (\ref{charrel}). The
characteristic function of a classical AGN (additive Gaussian noise, here assumed to be zero mean and circularly symmetric on the complex plane) random variable of variance $\ovl{n}$
is $\chi_{\ovl{n}}(\mu) = e^{-\ovl{n}|\mu|^2}$. In the Heisenberg
picture, AGN can be represented as
\beq \label{cagn} b=a+n \eeq
where $b$ is the output annihilation operator and $n$ is a c-number
complex-valued random variable with the above characteristic
function. From (\ref{charrel})-(\ref{cagn}) it follows that the addition of AGN with variance 1 is
sufficient to turn any state classical, which is a \emph{crucially
important} fact. \\ \\
\textbf{Lemma 1}: For an arbitrary input state, the output state after addition of Gaussian
noise of variance $1$ is  classical. \\ \\
\textbf{Proof}:  Using the fact that the $a$ and $n$ modes in (\ref{cagn}) are statistically independent, we find that $\chi_N^b(\mu) = \chi_N^a(\mu)\cdot e^{-|\mu|^2} = \chi_A^a(\mu)$ from (\ref{charrel}), so
that $P_b(\al)=Q_a(\al) \geq 0.\ \blacksquare$
\\ \\
\textbf{Lemma 2}: If a state has a nonnegative Wigner distribution $W(\al)$, the output state after addition of Gaussian noise of variance $1/2$ is classical. \\\\
\textbf{Proof}: In this case we have $P_b(\al)=W_a(\al) \geq 0$ by assumption. $\blacksquare$\\
\\
This result shows that for states with nonnegative $W(\al)$ only $1/2$ AGN noise photon per mode is sufficient to turn it classical. It is easily verified that multimode generalizations of lemmas 1 and 2 hold when statistically independent AGN of variance $1$ and $1/2$ respectively is added to each mode.

\section{Classicalization of Nonclassical States in Loss and Noise}

The usual linear loss is very detrimental to nonclassical states as stressed in the beginning for squeezed states \cite{yuen76}, the same being true for number states of large photon number and indeed any nonclassical states. This can be seen easily from the linear loss representation \cite{yuen76,yuen04} between the output annihilation operator $b$ and the input annihilation operator $a$ of a loss map with transmittance $\kappa$
\beq \label{lossmap}
b=\sqrt{\kappa}a+\sqrt{1-\kappa}c, \eeq
where the joint state $\rho_{ac}=\rho_a\otimes
\rho_c$ with $\rho_c = \ket{0}_c\bra{0}$, the vacuum state in the case of no added background noise. In (\ref{lossmap}), we have skipped the `$\otimes$' for $a=a\otimes I$ etc. on $\cl{H}_a\otimes\cl{H}_c$, as usual. For $\kappa \ll 1$, (\ref{lossmap}) shows that the mode $b$ is largely made up of the mode $c$ in vacuum which is a classical state. We will denote the linear loss CP-map by $\cl{L}_{\kappa}$, so that $\rho_b = \cl{L}_{\kappa} \rho_a$.

From (\ref{charfuncs}) and (\ref{lossmap}), the antinormally-ordered characteristic function
$\chi_A^b(\mu)$ of $\cl{L}_{\kappa} \rho$ is, for $\rho_a=\rho$ and $\rho_c = \ket{0}_c\bra{0}$ \beq \label{losschar}
\chi_A^b(\mu) = \chi_A^{a}(\sqrt{\kappa} \mu)e^{-(1-\kappa)|\mu|^2}.
\eeq
We use $\cl{G}_n$ to denote the CP-map corresponding to (\ref{cagn}) with  AGN of variance $n$. If an AGN of variance $\kappa$ is added to
$\cl{L}_{\kappa}\rho$, we have, for $\rho_o = \cl{G}_{\kappa}
\cl{L}_{\kappa} \rho,$
\beq \chi_N^{\rho_o}(\mu) =
\chi_A^{\rho_o}(\mu) e^{|\mu|^2} = \chi_A^{\rho} (\sqrt{\kappa}\mu), \eeq
where the second equality follows on using (\ref{losschar}).
In the multimode case, let $\rho$ be an arbitrary $m$-mode state, $\cl{L}_{\udl{\kappa}}$ the linear loss CP-map with transmittance $\udl{\kappa} = (\kappa_1, \ldots, \kappa_m)$, and $\cl{G}_{\udl{N}}$ the Additive Gaussian Noise (AGN) CP-map with noise variance  $\udl{N}=(N_1, \ldots, N_m)$, and again  $\rho_o = \cl{G}_{\udl{\kappa}}\cl{L}_{\udl{\kappa}} \rho$. The noise pdf is again assumed to be zero mean and circularly symmetric in each mode. We calculate as before that
\beq \label{multimodemap} \chi_N^{\rho_o}(\udl{\mu}) = \chi_A^{\rho}(\sqrt{\kappa_1}\mu_1, \cdots, \sqrt{\kappa_m} \mu_m). \eeq
The following theorem then asserts that in the presence of transmittance vector $\udl{\kappa}$, an AGN of variance $\udl{N}=\udl{\kappa}$ turns any $\rho$ classical, viz., \newpage\textbf{Theorem 1 (Classicalization Theorem)}:\\
\\
The state $\rho_o = \cl{G}_{\udl{\kappa}}\cl{L}_{\udl{\kappa}}\hspace{1mm} \rho$ is classical for any state $\rho$ and has the P-representation
\beq \label{thm1}
P_{\rho_o} (\udl{\al}) = (\prod_{i=1}^{m} \frac{1}{\kappa_i}) Q_{\rho} (\frac{\al_1}{\sqrt{\kappa_1}}, \ldots, \frac{\al_m}{\sqrt{\kappa_m}}).
\eeq\\
\\ \textbf{Proof}: Take the Fourier transform on both sides of (\ref{multimodemap}). $\blacksquare$
\\\\
Intuitively, a linear \textit{loss} $1-\kappa$ in each mode couples in an AGN of variance $N=1-\kappa$, which with the addition of a further AGN of variance $\kappa$ furnishes enough noise to turn the state classical, since any state becomes classical when an AGN $N=1$ corresponding to the heterodyne vacuum fluctuation level is added, as formalized in Lemma 1. Thus, any multimode $\rho$ after a large loss $\kappa_i \ll 1, i \in\{1,\ldots,m\}$ in each mode is nearly classical, differing from one such state $\rho_o = \cl{G}_{\udl{\kappa}}\cl{L}_{\udl{\kappa}}\hspace{1mm} \rho$ by an AGN map of variance only $\udl{\kappa}$. This fact \emph{is} the ghost that haunts the use of any novel quantum state in any situation with significant loss. In view of Lemma 2, only an AGN of variance $(\kappa/2)$ is needed when the input state $\rho$ has a nonnegative Wigner distribution.

The application of the Classicalization Theorem to a sensing problem that uses a nonclassical transmitter state $\rho$ is as follows. The sensor performance is determined by the received state $ \rho_\mrm{out}=\cl{L}_{\kappa}\rho$, considering for now only a pure loss channel $\cl{L}_{\kappa}$. Consider transmitting instead the classical state $\tilde{\rho} = \cl{G}_{1} \rho$ of Lemma 1. The received state in this case is
\beq \label{receivedclassicalizationstate}
\cl{L}_{\kappa}\tilde{\rho} =\cl{L}_{\kappa}\cl{G}_{1}{\rho} = \cl{G}_{\kappa}\cl{L}_{\kappa}\rho =\cl{G}_{\kappa} \rho_\mrm{out} \equiv \tilde{\rho}_\mrm{out}. \eeq
The penultimate equality follows from the relation $\cl{L}_{\kappa}\cl{G}_N = \cl{G}_{\kappa N} \cl{L}_{\kappa}$ which can be verified by following the Heisenberg picture mode transformations given in equations (\ref{cagn}) and (\ref{lossmap}). Thus the performance difference in the two cases is characterized by the difference between the density operators $\rho_\mrm{out}$ and $\tilde{\rho}_\mrm{out}=\cl{G}_{\kappa} \rho_\mrm{out}$. If $\kappa$ is small, the performance cannot be expected to be very different. In Section IV, the difference is quantified using the trace distance in sensors for which additional system noise makes $\rho_\mrm{out}$ classical.

\section{Classical versus Nonclassical State Comparison Via Trace Distance}

We are interested in delimiting the parameter regions in which nonclassical transmitted states could not lead to significant performance gain compared to classical states. In this section, strong quantitative results in the form of sufficient conditions are provided in the case of a \emph{single} transmitted ``signal'' mode with or without entanglement to an ``idler'' mode kept at the transmitter or receiver. These results are obtained by developing trace distance bounds on the two output states corresponding respectively to the nonclassical transmitted state and a classical one containing one added signal photon at the transmitter.

The trace distance bound can quantify the performance difference of the two states in general as follows. For any given Positive Operator Valued Measure (POVM) $X$ corresponding to a quantum measurement, let $X(\Delta)$ be the POVM element representing the
measurement result $x \in \Delta$ where $\Delta$ is a Borel set on
$\mathbb{R}^n$. The probability of such a result when the state is
$\rho$ is given by $P_{\rho}(x \in \Delta) = \tr \rho X(\Delta)$.
Since  \cite{schatten70}
\beq \label{holder}
|\tr \rho X| \leq \|\rho\|_1 \norm{X} \eeq
for any trace-class operator $\rho$
with trace norm $\norm{\rho}_1$ and bounded operator $X$ with the
usual operator
norm $\norm{X}$, we have \\
\\
\textbf{Lemma 3}:  Let two states $\rho_1, \rho_2$ have
$\norm{\rho_1 -\rho_2}_1 \leq \e$. Then the probability of any
measurement result $x \in \Delta$ differs at most by $\e$ for the
two states, i.e.,
\beq \label{probdif} |P_{\rho_1}(x \in \Delta) -
P_{\rho_2} (x \in \Delta)| \leq \e. \eeq
\\
\textbf{Proof}: As $P_{\rho_i}(x \in \Delta)= \tr \rho_i
X(\Delta), i=1,2$, (\ref{probdif}) follows from (\ref{holder}) since
$\norm{X(\Delta)} \leq 1$ for any POVM element $X(\Delta).\hspace{2mm}  \blacksquare$
\\ \\
It is easy to see that any performance criterion of a system in
state $\rho$ is a function of the probability $P_{\rho}(x \in
\Delta)$. For typical criteria which do not diverge hugely for
small differences in $P_{\rho}(x \in \Delta)$, a sufficiently small $\e$ in
(\ref{probdif}) would guarantee the practical equivalence of
$\rho_1$ and $\rho_2$.

Consider a nonclassical state $\rho_\mrm{in}$ which in loss $1-\kappa$ and AGN at the receiver leads to a received state $\rho_\mrm{out}$ which is classical. In our notation, $\rho_\mrm{out}= \cl{G}_N \cl{L}_{\kappa} \rho_\mrm{in}$ is classical. Such a classical $\rho_\mrm{out}$ is not in general obtainable from a classical transmitted state due to the fact that both $\cl{G}$ and $\cl{L}$ are one-to-one maps on the space of density operators. This follows from noting that the transformations of the characteristic functions given in lemma 1 and (\ref{losschar}) corresponding to $\cl{G}$ and $\cl{L}$ are both one-to-one, so that only one input state corresponds to a given output state (The mathematical invertibility of the maps does not of course imply that the inverse can be physically implemented). Since an AGN of variance $=\kappa$ turns the state $\cl{L}_{\kappa} \rho_\mrm{in}$ classical, we could compare $\tilde{\rho}_\mrm{out} =\cl{G}_\kappa \rho_\mrm{out}= \cl{G}_N\cl{G}_{\kappa}\cl{L}_\kappa \rho_\mrm{in}$  to $\rho_\mrm{out}$. The former can be obtained from a classical transmitted state $\tilde{\rho}_\mrm{in} = \cl{G}_1 \rho_\mrm{in}$ as mentioned before. These two input states are related by
\beq \label{classicalizationstate}
P_{\tilde{\rho}_\mrm{in}} (\al) = Q_{\rho_\mrm{in}} (\al) \eeq
from the classicalization theorem. From (\ref{classicalizationstate}), the average photon number in $\tilde{\rho}_\mathrm{in}$ is greater than that in $\rho_\mrm{in}$ by one photon per mode
\beq \label{energyincrease}
\tr \tilde{\rho}_\mrm{in} a^{\dag} a = 1+ \tr {\rho}_\mrm{in} a^{\dag} a.\eeq
If the operating frequency is below X-ray frequency, one photon per mode corresponds to a tiny power that is negligible though we will return to this issue later. Assuming for now that the replacement of $\rho_\mrm{in}$ by $\tilde{\rho}_\mrm{in}$ is justified, we must now compare $\rho_\mrm{out}$ to $\tilde{\rho}_\mrm{out}$. This is achieved by \\ \\
\textbf{Theorem 2}:
 Let $\rho$ be a single-mode classical state. Then
 \beq \label{outputtracedistancebound}
 \norm{\rho - \cl{G}_N \rho}_1 \leq 2[N/(N+1)]^{1/2}. \eeq
 \\
 \textbf{Proof}:  We have the following triangle
 inequality for any norm $\norm{\cdot}$ and a scalar function $f(\lambda)$,
 \beq \label{intnorm}
 \|\int f(\lambda) A(\lambda) d\lambda\hspace{3mm}  \|  \leq \int
 |f(\lambda)| \norm{A(\lambda)} d\lambda, \eeq
 which follows from the integral as the limit of an infinite sum
 and the preservation of inequality under limit. From (\ref{classical}) and (\ref{intnorm}) we have
 \beq \norm{\rho - \cl{G}_N
 \rho}_1 \leq \int P(\al) \norm{\ket{\al}\bra{\al} - \cl{G}_N
 (\ket{\al}\bra{\al})}_1 d^2\al.
  \eeq
From the Uhlmann upper bound \cite{uhlmann76} on trace distance we have
\beq \norm{\ket{\al}\bra{\al} - \cl{G}_N
 (\ket{\al}\bra{\al})}_1  \leq 2[1- \bra{\al}\rho'\ket{\al}]^{1/2},
                                \eeq
where $\rho' = \cl{G}_N (\ket{\al}\bra{\al})$ has $Q(\beta)$ given by \cite{ks68}
 \beq Q_{\rho'}(\beta)= \frac {1}{\pi(N+1)} e^{- \frac {|\beta
 -\al|^2}{N+1}}. \eeq
 The bound (\ref{outputtracedistancebound}) follows from (\ref{intnorm})-(17). $\blacksquare$
\\ \\
It is clear from (\ref{probdif}) and (\ref{outputtracedistancebound}) that there can be no significant advantage with single-mode nonclassical states for $N=\kappa \ll 1$.

It is also possible to compare the case where an AGN of variance $N$ at the receiver turns the state $\cl{L}_{\kappa} \rho_\mrm{in}$ classical already. In such a situation, we would be comparing $\cl{G}_{N} \rho$ and $\cl{G}_{N+\kappa} \rho$ for $\rho = \cl{L}_\kappa \rho_\mrm{in}$. For the case $N$ is not small, we may use the following \\ \\
\textbf{Theorem 3}: For a classical state $\rho$, \beq \label{anothertracedistancebound}
\norm{(\cl{G}_{N_1}- \cl{G}_{N_2})\rho}_1 \leq \frac
{2(N_1-N_2)} {N_2} \eeq where we take $N_1
\geq N_2$. \\ \\
\textbf{Proof}: For $\rho$ with P-representation $P(\al)$, \beq
\norm{(\cl{G}_{N_1}- \cl{G}_{N_2})\rho}_1 = \int {d^2
\al P(\al)(\cl{G}_{N_1}-
\cl{G}_{N_2})(\ket{\al}\bra{\al}) } \eeq and so from (\ref{intnorm})
\beq \norm{(\cl{G}_{N_1}- \cl{G}_{N_2})\rho}_1 \leq
\int {d^2 \al P(\al)\norm{(\cl{G}_{N_1}-
\cl{G}_{N_2})(\ket{\al}\bra{\al})}_1 }. \eeq From (\ref{intnorm}) and
(\ref{classical}), \beq \norm{(\cl{G}_{N_1}-
\cl{G}_{N_2})(\ket{\al}\bra{\al})}_1 \leq \int{
\frac{d^2\beta}{\pi}|\frac{1}{N_1} e^{- \frac{|\beta
-\al|^2}{N_1}} - \frac{1}{N_2} e^{- \frac{|\beta
-\al|^2}{N_2}} | }. \eeq By writing $1/{N_1} =
1/{N_2} - \delta$ so that $\delta = \frac {N_1 -
N_2} {N_1 N_2}$, the right hand side of (21) is
itself less than \beq \int{ \frac{d^2\beta}{\pi N_2}\{ e^{-
\frac{|\beta -\al|^2}{N_1}} -  e^{- \frac{|\beta
-\al|^2}{N_2}} }\} + \delta \int{ \frac{d^2\beta}{\pi} e^{-
\frac{|\beta -\al|^2}{N_1}} }. \eeq The bound (\ref{anothertracedistancebound}) follows
from evaluation of the integrals in (22). \ $\blacksquare$
\\ \\
For the two-mode case, similar to the derivation in Theorem 3, we have the following\\ \\
\textbf{Theorem 4}: Let $\rho$ be a two-mode classical state and $\udl{N}^a= (N^a_{1}, N^a_{2})$ and $\udl{N}^b= (N^b_{1}, N^b_{2})$ be two-mode noise variance vectors with $\udl{N}^a \geq \udl{N}^b$ componentwise. We then have \beq
\label{twomodetracedistancebound} ||(\cl{G}_{\udl{N}^a} - \cl{G}_{\udl{N}^b}) \rho||_1  \leq   2[ \frac {N^a_1 - N^a_2} {N^a_2} + \frac{N^b_1 - N^b_2}
{N^b_2} + 2\frac{(N^a_1 - N^a_2)(N^b_1
- N^b_2)} {N^a_2 N^b_2}]. \eeq
\\
Consider the case of a two-mode entangled sensor where the idler mode $2$ is assumed perfectly preserved but there is loss and noise in the signal mode $1$ so that $\udl{\kappa} \sim (\kappa,1), \kappa \ll 1$ and $\udl{N} = (N,0)$. When the total received state is classical, the bound (\ref{twomodetracedistancebound})
yields the difference between $\rho_{out}$ and $\tilde{\rho}_{out}$ via \beq \norm{\cl{G}_{\udl{N}} \cl{L}_{\udl{\kappa}}
(\rho_\mrm{in} - \tilde{\rho}_\mrm{in}) }_1 \lesssim 2\kappa/N, \eeq
which shows that there can be no significant performance
improvement for any signal level.

In the multimode case, one may expect the trace distance to
 increase as the states become more distinguishable. The following
 multimode generalization of (\ref{outputtracedistancebound}) can be obtained via \\
 \\
 \textbf{Lemma 4}: For any operators $A,B,C,D$ and the operator norm $\parallel \cdot \parallel$,
 \beq \label{multioperatorinequality} \norm{A\otimes B - C \otimes D}_1 \leq \norm{B}
 \norm{A-C}_1 + \norm{C} \norm{B-D}_1. \eeq
 \textbf{Proof}: From
 $A\otimes B -C\otimes D = (A-C)\otimes B + C \otimes (B-D)$, we
 have (\ref{multioperatorinequality}) from the triangle inequality and (\ref{holder}) where
 $\norm{\cdot}$ is the usual operator norm $\blacksquare$.
 \\ \\
If we apply (\ref{multioperatorinequality}) to $\norm{\ket{\al}\bra{\al} -
\cl{G}_N
 (\ket{\al}\bra{\al})}_1$ and use the bound $\norm{\rho} \leq 1$
 for any state $\rho$, we obtain \\
 \\
 \textbf{Theorem 5}:  Let $\rho$ be an $m$-mode classical state. Then
 \beq \label{multimodetracedistancebound}
 \| \rho - \cl{G}_{\udl{N}}\rho \|_1 \leq 2\sum_{i=1}^{m}{[\frac{N_i}{{N_i+1}}]^{1/2}},
\eeq
 where $\cl{G}_{\udl{N}}$ is the multimode AGN map with vector variance $\udl{N}$.\\ \\
 Note that the trace distance between any two states satisfies $0 \leq \|\rho_1 -\rho_2\|_1 \leq
 2$, 0 when $\rho_1= \rho_2$ and 2 when $\rho_1$ and $\rho_2$ have
 orthogonal ranges. Thus, the bound (\ref{multimodetracedistancebound}) is useful in the case $N_i \ll
 1$ only when the sum on the righthand side  is less than 1. For ${N}_i \sim
 N \ll 1$, this is equivalent to a ``bandwidth'' condition $m \lesssim N^{-1/2}$. Thus,
 Theorem 5 has the following implication for sensors where
 $\kappa_i < \kappa \ll 1$ and an AGN of $N_i \geq \kappa$ is
 present: For $m < \kappa^{-1/2}$, one cannot expect any significant
 improvement over classical states by transmitting quantum states.
 This is because if the classical state satisfying (\ref{classicalizationstate}) is
 transmitted instead, the output $\cl{G}_{\ovl{N}}\rho$ is
 little different from the output $\rho$ using whatever quantum
 state from (26).

One may expect that there may not be a generally useful tight trace distance upper bound in the multimode case as follows. Even without entanglement, the Hilbert-Schmidt inner product of two multimode states $\otimes_i \rho_i$ and $\otimes_i \sigma_i$ is $\prod_i \tr \rho_i \sigma_i$, which tends to zero for a large number of modes since $\tr \rho \sigma <1$ for any $\rho \neq \sigma$. Thus a small difference between $\rho$ and $\sigma$ could be magnified arbitrarily by increasing the number of modes as the total states become asymptotically orthogonal at least in the pure state case. In such a case, $\|{\otimes}_i \rho - \otimes_i \sigma\|_2 \rightarrow 2$. From the inequality $\| A\|_1 \geq \|A\|_2,$ it follows that the trace distance also tends to its maximum value $2$. However, this kind of ``law of large numbers'' effect also occurs classically and is clearly not the kind of performance improvement we are looking for from nonclassical states, e.g., the large improvement possible with squeezed states for just a single mode. On the other hand, the possibility of multimode entanglement complicates the situation and there is much to be learnt and quantified on the problem of classical versus nonclassical state comparison. In the following, we present a different type of No-Go argument via the notion of a ``practical equivalence class'' of states that is applicable to the multimode situation.

\section{Practical No-Go Theorem}
We employed the following no-go strategy in section IV. If the system output suffers an AGN at least as large as $\udl{\kappa}$, then it is a classical state given by (1) which can be obtained from a transmitted classical state with one more photon per mode due to the change from $Q(\cdot)$ to $P(\cdot)$ in (\ref{thm1}). We asserted that this is an insignificant energy difference. The output of such a classical transmitted state differs from the output of an input $\rho$ by just an AGN of variance $\udl{\kappa}$. We bounded the trace distance between these two outputs which provides a universal quantitative bound on the probabilities obtainable from any measurement on the two different states. The results given in Theorems 3-5 are not strong enough to cover all relevant multimode possibilities. This is due to the possibility of accumulating small advantages over many modes to get a large one in an essentially classical way.

However, for $\kappa \ll 1$, say $\kappa \sim 10^{-10}$, it is intuitively clear that an AGN of variance $\udl{\kappa}$ has no ``practical meaning'' in that the system model is never accurate enough to make a noise as small as the fraction $\kappa/(1-\kappa) \sim \kappa$ relative to the vacuum fluctuation meaningful. States differing only by $\cl{G}_{\udl{\kappa}}$ for such small $\udl{\kappa}$ can then not be experimentally distinguished and thus fall in the same ``practical equivalence class''. Perhaps equally important is that it is difficult to give a meaningful performance criterion that would depend on such a small noise difference. Indeed, if such a situation is found to be practically important, one must obtain the corresponding mathematical system model to such a high accuracy \emph{first}. Thus we find the following argument sufficient for No-Go with quantum sensors in the region $\kappa_i \ll 1$ for all the modes.

\begin{itemize}
\item (\textbf{A}) An increase of one photon per mode at the transmitter can be seen to be a practically irrelevantly small amount at optical frequencies and below for any one of the following reasons:
\begin{indent}
\begin{enumerate}[]
\item (\textbf{A1}) \begin{indent} It translates to a power of $-10$ dBm at $\lambda \sim 1\hspace{1mm} \mu m$ even for the maximum bandwidth of $W \sim 10^{15}$ Hz and is thus easily obtained from laser sources.\end{indent}
\item (\textbf{A2}) The amplitude fluctuation of a laser also by far exceeds one photon per mode for easily available moderate source power.
\item (\textbf{A3}) It is hard to envisage scenarios in which one needs to exclude such small increases in source power (although they cannot be absolutely ruled out, as discussed below). Indeed, in a meaningful classical versus quantum comparison, the transmitter energy or power constraint itself is expressed by
    \beq \label{absoluteenergyconstraint}
    N_S \leq N_\mrm{max} \textrm{\hspace{4mm}or\hspace{4mm}} P_S \leq P_\mrm{max} \eeq
    for \emph{per mode} signal energy $N_S$ or \emph{total} power $P_S$, where the $N_\mrm{max}$ and $P_\mrm{max}$ are determined by what can readily be achieved in practice and can be tolerated for the given application. Artificial restriction to smaller values would not reflect the true classical capability available.
\end{enumerate} \end{indent}
\item (\textbf{B}) An AGN of variance $\kappa \ll 1$ is practically irrelevantly small because it is a very small fraction of the vacuum fluctuation noise $1-\kappa$ introduced by way of the channel loss. The system model does not have such accuracy in practice and the system  performance is not expected to vary markedly under such a small change in the noise variance.
\end{itemize}

To elaborate and justify these points, we note regarding (A1) that the average power $P_S$ and pulse duration $\tau$ are related via $P_S=\hbar\omega_0 N_S/ \tau$, for $N_S$ the average photon number per pulse. The total number of modes is $M = TW$, where $W$ is the signal bandwidth and $T> \tau$ is the total signaling time. For the best possible case of transform-limited pulses with $\tau W \sim 1$, we have
\beq \label{power}
P_S =\hbar \omega_0 N_S W. \eeq
Note that the bandwidth $W$ satisfies $W < \omega_0$ for  carrier frequency $\omega_0$ and that the total number of pulses in time $T$ is $T/\tau = M$. For $N_s \sim 1$, this gives a very small power at the transmitter at optical frequencies and below, but could be considerable at much higher frequencies. At $\lambda \sim 1 \mu\textrm{m}$, $1\hspace{1mm} W$ of power corresponds to $\sim 10^{19}$ photons per sec. With picosecond pulses, $N_s \sim 1$ corresponds to $-40$ dBm while a diode laser used for optical fiber transmission already puts out $0$ dBm or $1$ mW. Clearly, this one-photon power is thus a tiny fraction of what readily available laser sources give. Indeed, the intensity fluctuation of such a source, i.e., the excess noise in the parameter $|\alpha|^2$ over and above the coherent state quantum fluctuation, is typically a few percent, which is many times this small power -- this is our point (A2). Note also that for given $P$, $N_S$ cannot be made smaller than what the bandwidth $W$ allows, from (\ref{power}). Note also that at X-ray frequencies and above, one photon per mode starts to become a significant power so that (A) may not hold anymore.

 Our point (A3) states that the energy or power constraint should be given by what is \emph{readily available} from a classical-state source and \emph{readily tolerated} in the application of interest, and $N_S$ and $P_S$ should not be restricted to an artificially small range. It is important to note that it is source power that one is concerned with in typical applications, and other quantities of interest, including the total energy, scale proportionately with the power for a fixed time interval $T$.
If one extends the total energy from small $N_s$ in (\ref{power}) with nonclassical source and long time interval $T$, one should compare it with a classical state for the same $T$ but an $N_\mrm{max}$ of (\ref{absoluteenergyconstraint}) that is much bigger as long as this extra energy is readily available and readily tolerated by the system. That said, the numerical values of $N_\mrm{max}$ or $P_\mrm{max}$ depend on the application. It is conceivable that there may be scenarios where $N_\mrm{max}$ and $P_\mrm{max}$ need to be smaller than the increases stipulated by our classicalization theorem either for tactical or physical reasons. We exclude such scenarios from the scope of our No-Go theorem.

In view of these considerations, we conclude that at optical frequencies and below, the difference of one photon per mode at the transmitter that is needed for the equivalent input classical state $\tilde{\rho}_\mrm{in}$ compared to $\rho_\mrm{in}$ is not relevant in typical practice. Indeed, we would argue that it is not relevant in principle either so long as $N_\mrm{max} \gg 1$ in (\ref{absoluteenergyconstraint}). Physically one cannot give a precise meaning to such a small fraction in reality, and indeed the situation \emph{requires} a fundamentally more precise description if it is sensitive to such small changes.

A similar consideration applies  to point (B). It is hardly ``physically'' meaningful to consider an AGN of variance $\kappa \ll 1$ to be relevant in the presence of another AGN of variance $1-\kappa$, from either a practical point of view or from that of a mathematical model of reality. As a consequence of (A) and (B), we argue: \\ \\
\textbf{Practical NO-GO on Nonclassical Transmitter in Loss}: \\
\\ If $\kappa_i \ll 1$ for all $i$, there is no practical difference between $\cl{L}_{\udl{\kappa}}\hspace{1mm} \rho$ and $\cl{G}_{\udl{\kappa}}\cl{L}_{\udl{\kappa}}\hspace{1mm} \rho$, the latter being obtainable from a classical transmitted state which has no practical difference in average power from that of $\rho$.\\
\\
Of course, the problem is, exactly speaking, quantitative. When $\kappa$ is not too small, say $\kappa \sim 10^{-1}$, the point (B) above gets shaky while it is clearly valid for $\kappa \sim 10^{-10}$. However, if there is already some AGN with $\udl{N} \sim \udl{\kappa}$ in the system, the argument is sufficiently strong for any value of $\udl{\kappa}$ because the received state is classical for $\udl{N} \sim \udl{\kappa}$. Even in the absence of precise bounds as provided under more restrictive conditions in section IV, the AGN required for classicalization is just too small to make any difference with the vacuum fluctuations from loss already present. In fact, the above No-Go is \emph{quantitative} in that one can judge in any particular problem how negligible AGN of variance $\kappa$ is in the background of another AGN of variance $1-\kappa$.

In the case of entangled transmitter states where the kept idler modes are assumed to suffer no noise or loss, the above No-Go theorem does not apply. If the total received state turns out to be classical in such a case, it is because the noise added to the receiver signal mode is sufficient to make the total signal plus idler state classical. This occurs in the case of the ``quantum illumination'' Gaussian state \cite{tan08}, specifically for the downconverter state that we call QI-DC here, where
\beq \label{qidc}
\ket{\psi}=\sqrt{1-|\lambda|^2}\sum_{n} \lambda^n \ket{n}_a\ket{n}_b, \hspace{3mm} \lambda=\sqrt{\frac{N_s}{N_s+1}}.
\eeq
In this case, it turns out that the addition of a \emph{single} AGN photon to the signal mode \emph{alone} is actually sufficient to turn the total signal plus idler state classical. Since the state (\ref{qidc}) is Gaussian, the classicality of the state with one added signal AGN photon may be verified by checking that its Wigner Covariance matrix is positive semidefinite.

For this entangled transmitter state scenario where the total received state is classical, a No-Go can be formulated even though the idler $\kappa_i \sim 1$. Because one added photon in the form of AGN to the signal mode makes the state (31) classical, the output received state for this classical input state is identical to  the received state from the entangled transmitted state (\ref{qidc})) except for an additional AGN of variance $= \kappa_s$, the signal mode transmittance. Thus the points (A) and (B) of this section imply a similar No-Go for the state (\ref{qidc}) for $\kappa_s \ll 1$.

This No-Go bears directly on the results of \cite{lloyd08,tan08}. The performance gain in \cite{lloyd08} obtained under rather general quantum illumination states is predicated on single photon detection which is not a realistic limitation for a receiver, as pointed out in \cite{shapiro09}. More remarkable performance gain is reported in \cite{tan08} for QI-DC in the presence of large signal-mode receiver noise $N_B \gg 1$ but small signal mode energy $N_s \ll 1$ per mode, as compared to the performance of classical transmitted states of the same $N_s$. Since Theorem $4$ or (\ref{twomodetracedistancebound}) precludes such performance gain for a single pair of (\ref{qidc}), the performance advantage is obtained in the multimode situation. On the other hand, theorem 5 leads to a useless bound (28) since the right-hand side is bigger than $2$ for the case of large $m > 10^5$ and  $N_i \sim 20$ in the signal modes.

The QI-DC performance is surprising, and is perhaps indicative of the power of a multimode quantum receiver, but is not totally surprising because $N_s \ll 1$ is really a ``microscopic'' quantum limit, and the advantage of QI-DC disappears as $N_s$ increases beyond $1$. For a fixed $N_s$, arbitrary advantage cannot be gained by increasing the total energy $N=M N_s$ because the available mode number $M$ is limited by the bandwidth.  Our Practical No-Go says that there is no advantage if the power corresponding to $N_s=1$ per mode is negligible, which we have argued is indeed the case for various reasons for realistic classical-state sources. The key reason that no-go applies in the case of quantum illumination is that AGN of small variance at the signal mode alone turns the transmitted state classical. In a practical scenario where the power (\ref{power}) corresponding to $N_s \sim 1$ is required, perhaps because one wants to minimize the power hitting a target, restriction to such small $N_s$  is difficult to reconcile with the fact that the  QI-DC advantage already requires $N_B \gg 1$. Indeed, it seems to us that the small numerical value of (\ref{power}) for $N_s \sim 1$ at optical frequencies and below suggests that it could not make any real difference at the transmitter and much less so at the target due to the large attenuation by $\kappa$. However, we leave open the possible advantages of QI-DC in scenarios where $N_\mrm{max}$ is constrained to be small.

\section{Conclusion}
We have developed some rigorous quantitative bounds and density operator representations that differentiate the performance that one may obtain by employing nonclassical or entangled transmitter states as compared to classical ones. The No-Go results presented in Section IV are quite general and criterion-independent, except that they depend on the assumption that the total received state is classical. In Section V, a practical No-Go theorem is enunciated that holds when there is large loss in all modes. A similar result covers the scenarios involving the recent quantum illumination states. The general case of nonclassical transmitted states for moderate loss and noise is very significant and is yet to be investigated thoroughly. This is especially so for true multimode entanglement that is not just a statistical accumulation of entanglements between two modes.

\section{Acknowledgements}
We would like to thank Prem Kumar and Jeffrey H.~Shapiro for useful discussions. This work was supported by the DARPA Quantum Sensors Program.


\begin{thebibliography}{10}
\expandafter\ifx\csname url\endcsname\relax
  \def\url#1{\texttt{#1}}\fi
\expandafter\ifx\csname
urlprefix\endcsname\relax\def\urlprefix{URL }\fi


\bibitem{yuen76}
H.P.~Yuen,  Phys. Rev. A \textbf{13}, 2226 (1976).

\bibitem{yuen04}
H.P.~Yuen, in \emph{Quantum Squeezing}, ed. by P.D.~Drummond and Z.~Ficek, Springer Verlag (2004), pp.~227-261; also quant-ph/0109054.

\bibitem {lloyd08}
S.~Lloyd, Science  \textbf{321}, 1463 (2008).

\bibitem{tan08}
S.-H.~Tan, B.I.~Erkmen, V.~Giovannetti, S.~Guha, S.~Lloyd, L.~Maccone, S.~Pirandola, and J.H.~Shapiro, Phys. Rev. Lett. \textbf{101}, 253601 (2008).

\bibitem{schatten70}
R.~Schatten, \emph{Norm Ideals of Completely Continuous Operators}, Springer, 1970, p.~39.

\bibitem{uhlmann76}
A.~Uhlmann, Rep. Math. Phys., 9, 273 (1976).

\bibitem{ks68}
J.R.~Klauder and E.C.G.~Sudarshan, \emph{Fundamentals of Quantum
Optics}, Benjamin, 1968 (reprinted by Dover 2006).

\bibitem{shapiro09} J.H.~Shapiro and S.~Lloyd, preprint submitted to New J. Phys.; also arXiv:0902.0986.

\end{thebibliography}
\end{document}